\documentclass[12pt]{article}
\usepackage{graphicx,amsmath,amssymb,dcolumn,afterpage,supertabular,epsfig}
\usepackage[hang,bf]{caption}
\usepackage{thumbpdf}
\usepackage[colorlinks=true, linkcolor=blue, citecolor=blue, urlcolor=blue,
pdfauthor={Philip Bechtle, Klaus Desch, Werner Porod, Peter Wienemann},
pdftitle={Determination of MSSM Parameters from LHC and ILC 
Observables in a Global Fit},
pdfkeywords={SUSY SPA Fittino}]{hyperref}

\newcolumntype{d}[1]{D{.}{.}{#1}}

\begin{document}

\thispagestyle{empty}

\begin{flushright}
IFIC/05-39 \\
ZU-TH 16/05 \\
SLAC-PUB-11580\\
\end{flushright}
\vspace{5mm}
 
\begin{center}
{{\LARGE \bf Determination of MSSM Parameters\\[1ex] from LHC and ILC Observables\\[2ex]
             in a Global Fit}\\[14ex]
{\sc Philip Bechtle$^{1}$, Klaus Desch$^{2}$, Werner Porod$^{3,4}$, Peter Wienemann$^{2}$}\\[3ex]}
{\sl $^1$Stanford Linear Accelerator Center (SLAC),\\
         2575 Sand Hill Road, Menlo Park, CA 94025, USA\\[1.5ex]
     $^2$Universit\"at Freiburg, Physikalisches Institut,\\
	 Hermann-Herder-Str.~3,
	 D-79104 Freiburg, Germany\\[1.5ex]
     $^3$Instituto de F\'isica Corpuscular, Universitat de Val\`encia\\
         Apartado de Correos 22085, E-46071 Val\`encia, Spain\\[1.5ex]
     $^4$Universit\"at Z\"urich, Institut f\"ur Theoretische Physik,\\
         Winterthurer Str.~190, CH-8057 Z\"urich, Switzerland
}\end{center}
 
\vspace{5mm}
 
\begin{abstract}
\noindent
We present the results of a realistic global fit of the Lagrangian
parameters of the Minimal Supersymmetric Standard Model assuming
universality for the first and second generation and real parameters.
No assumptions on the SUSY breaking mechanism are made. The fit is
performed using the precision of future mass measurements of
superpartners at the LHC and mass and polarized topological
cross-section measurements at the ILC. Higher order radiative
corrections are accounted for whereever possible to date.  Results are
obtained for a modified SPS1a MSSM benchmark scenario but they were
checked not to depend critically on this assumption. Exploiting a
simulated annealing algorithm, a stable result is obtained without any
{\it a priori} assumptions on the values of the fit parameters. Most of the
Lagrangian parameters can be extracted at the percent level or better
if theoretical uncertainties are neglected. Neither LHC nor ILC
measurements alone will be sufficient to obtain a stable result. The
effects of theoretical uncertainties arising from unknown higher-order
corrections and parametric uncertainties are examined
qualitatively. They appear to be relevant and the result motivates
further precision calculations.  The obtained parameters at the electroweak
scale are used for a  fit of the parameters at high energy scales within
the bottom-up approach. In this way regularities at these scales are explored
and the underlying model can be determined with hardly any theoretical
bias. Fits of high-scale parameters to combined
LHC$+$ILC measurements within the mSUGRA framework reveal that even
tiny distortions in the low-energy mass spectrum already lead to
inacceptable $\chi^2$ values. This does not hold for
``LHC only'' inputs.
\end{abstract}



\section{Introduction}

Provided low-energy Supersymmetry (SUSY)~\cite{susy} is realized in
Nature, the next generation of colliders, the Large Hadron Collider
(LHC)~\cite{lhc} and the International Linear Collider
(ILC)~\cite{ilc} are likely to copiously produce SUSY particles and
will allow for precise measurements of their properties. Once SUSY is
established experimentally, it is the main task to explore the unknown
mechanism of SUSY breaking (SSB).  While specific models of SUSY
breaking, like e.~g.~minimal supergravity (mSUGRA)~\cite{msugra} can
be tested against the data in a relative straight-forward manner (see
e.~g.~\cite{atlastdr}), an exploration of the parameters of the
general Minimal Supersymmetric Standard Model (MSSM)~\cite{mssm}
parameter space is significantly more ambitious. A first step in
this direction has been done in \cite{msugrafits} where a fit of
the real and flavour-diagonal MSSM parameters has been
performed using MINUIT \cite{ref:Minuit}. The use of MINUIT implied
some restrictions, e.~g.~the starting point for the fit has to be close
to the real values and also the correlation matrix depends
significantly on the starting point due to the large number of
parameters.  These limitations can be overcome by the methods
presented in \cite{ref:FittinoProgram} and it is among the aims
of this paper to demonstrate this in detail.

The general MSSM Lagrangian contains more than 100 new parameters
which are only related to each other through the unknown SSB
mechanism.  Many of these new parameters are CP-violating complex
phases which are to some extent limited by the absence of neutron and
electron electric dipole moments~\cite{edm} and flavour-nondiagonal
couplings which are bounded by the absence of flavour-changing neutral
currents both in the quark and lepton
sector~\cite{Eidelman:2004wy,Gabbiani:1996hi}. It seems therefore
justified to initially consider a somewhat constrained MSSM with real
parameters and flavour-diagonal couplings. Furthermore universality of
the first and second generation parameters appears to be a reasonable
approximation while large Yukawa couplings lead to significant
differences for the third generation. Applying these constraints, the
number of free parameters is reduced to 19, including the top quark
mass as a parameter to account for parametric uncertainties. While
still many, these parameters may be confronted with an even larger
number of independent observables at the LHC and ILC.  Thus it is of
high interest how well this 19-dimensional parameter space can be
restricted with future measurements.

The experimental collabarations ATLAS~\cite{atlas} and CMS~\cite{CMS}
at the LHC have performed detailed simulations of the possibilities to
extract mass information from their future data, predominantly from
the reconstruction of kinematic end-points of involving leptons and
jets in cascade decays of gluinos and squarks~\cite{ref:LHCILC}
produced in proton proton collisions at 14~TeV centre-of-mass energy.

At a future electron positron collider for collision energies up to
1~TeV, the International Linear Collider (ILC), the kinematically
accessible part of the superpartner spectrum can be studied in great
detail due to favourable background conditions and the well-known
initial state~\cite{ref:LHCILC}.

The attempt of an evaluation of the MSSM Lagrangian parameters and
their associated errors is only useful if the experimental errors of
the future measurements are known and under control. The by far
best-studied SUSY scenario to serve as a basis for such an evaluation
is an mSUGRA-inspired benchmark scenario, the SPS1a scenario.  For
this scenario with a relatively light superpartner spectrum, a wealth
of experimental simulations exists and has recently been compiled in
the framework of the international LHC/ILC study
group~\cite{ref:LHCILC}. In this paper we take a point close in
parameter space which is consistent with low energy data and
respects the dark matter constraints. We perform a global fit of the
19 parameters to those expected measurements augmented by
possible measurements of topological cross-sections (i.~e.~cross
sections times branching fractions) that will be possible at the ILC with
polarised beams. Since a detailed experimental simulation for some
measurements is lacking, we estimate their uncertainties
conservatively from the predicted cross-sections and transferring
experimental efficiency from well-studied cases.

While at leading order, certain subsets of the parameters of the
Lagrangian only influence certain subsets of observables (e.~g.~only
three parameters determine the masses of the charginos) this
'block-wise diagonal' mapping of parameters to observables no longer
holds at loop-level. At loop-level, in principle each observable
depends on each Lagrangian parameter, thus making the extraction of
the parameters much more involved.  This is particularly striking in
the supersymmetric Higgs sector where due to the large third
generation Yukawa couplings, radiative corrections to the mass of the
lightest Higgs boson are in general of the order of 30~\% to
50~\%. However, also for the superpartner properties radiative
corrections become important as soon as experimental precision enters
the percent level which is clearly the case for many of the
measurements possible at the LHC and in particular at the ILC.

The calculation of higher-order corrections to SUSY observables has
started in many sectors, see e.~g.~ref.~\cite{susycorrections} for an
overview. However, a coherent
framework for these corrections and a study of the transformation of
the parameters defined in different renormalisation schemes has only
started recently in the framework of the Supersymmetry Parameter
Analysis (SPA) project \cite{ref:SPAPaper}.

As theoretical basis for the global fit presented in this paper we use
the calculations as implemented into the program
SPheno~\cite{ref:SPheno}.  In SPheno, the masses and decay branching
fractions of the superpartners are calculated as well as production
cross sections in $\text{e}^+\text{e}^-$ collisions. The calculation
of the masses is carried out in the $\overline{DR}$ scheme and the
formulae for the 1-loop masses are used as given in
ref.~\cite{Pierce:1996zz}. In the case of the Higgs boson masses, the
2-loop corrections as given in ref.~\cite{Degrassi:2001yf} are
added. The calculation of the branching ratios is performed at
tree-level using however running couplings evolved at the scale
corresponding to the mass of the decaying particle. In case of the
cross sections tree-level formulae are used except for the production
of squarks in $\text{e}^+ \text{e}^-$ annihilation where the formulae
of ref.~\cite{Eberl:1996wa} are used. In addition we have added ISR
corrections as given in \cite{Kuraev:1985hb}. For the evolution of
parameters between various energy scales the 2-loop RGEs as given in
\cite{Martin:1993zk} are used.

In this paper we present the result of a global fit of the MSSM
Lagrangian parameters at the electro-weak scale for a slightly
modified SPS1a benchmark scenario using the program
Fittino~\cite{ref:Fittino,ref:FittinoProgram}. For a similar program
see \cite{sfitter}. Previous evaluations of the errors of those
parameters~\cite{msugrafits} did not attempt to develop a strategy to
extract those parameters from data without {\it a priori}
knowledge. Within Fittino, special attention is given precisely to
this task, i.~e.~to find the parameter set which is most consistent
with the data before a careful evaluation of errors and correlations
is performed. The obtained parameters are then evolved to high energy
scales, e.~g.~a GUT scale, taking into account all correlations between
the errors. This is done within the bottom-up approach, e.~g.~no 
assumptions regarding the underlying high energy model are used.
For further details see \cite{msugrafits}.

This paper is organized as follows. In Section~\ref{sec:fitprocedure}
we shortly describe the approach used in Fittino. In
Section~\ref{sec:SPS1aPrimeFit}, the modified SPS1a benchmark scenario
(SPS1a') and assumptions for the input observables are explained. The
results of the fit and the error evaluation method are summarized in
Section~\ref{sec:UnsmearedFitResultsSPS1aPrime}. The extrapolation of
the obtained fit parameters and their errors to high energy scales is
carried out in Section~\ref{sec:extrapolation}. A fit within the
mSUGRA framework is perfromed in Section~\ref{sec:mSUGRAFit} and
conclusions are drawn in Section~\ref{sec:conclusions}.


\section{Fit Procedure}
\label{sec:fitprocedure}

From the numerous fitting options provided by Fittino we have chosen
the following fit procedure to extract the low-energy Lagrangian
parameters. First start values for the parameters are calculated using
tree-level relations between parameters and a few
observables~\cite{ref:FittinoProgram}. For fits with many parameters
these values are not good enough to allow a fitting tool like
MINUIT~\cite{ref:Minuit} to find the global minimum due to the amount
of loop-level induced cross-dependencies between the individual
sectors of the MSSM. Therefore, in a second step, the parameters are
refined, using a simulated annealing
approach~\cite{ref:FittinoProgram,SimAnn,SimAnn2}. As a result, the
parameter values are close to the global minimum so that a global fit
using MINUIT can find the exact minimum in a third step. To determine
the parameter uncertainties and correlations many individual fits with
input values randomly smeared around their true values within their
uncertainty range are carried out. The parameter uncertainties and the
correlation matrix are derived from the spread of the fitted parameter
values.

The advantages of this approach are:
\begin{itemize}
    \item All available measurements can be used at once to extract
      the maximal possible information from the data.
    \item No {\it a priori} knowledge is required for the fit. The
      calculation of tree-level start values is a time-saving way to
      obtain reasonable start values for the fit. A brute force scan
      of the parameter space would not be feasible for a large set
      of parameters.
    \item Correlations between input observables can easily be taken
      into account.
\end{itemize}

The results presented in this article were obtained with Fittino
version 1.1.1. More detailed information on the algorithms
used by Fittino can be found in~\cite{ref:FittinoProgram}.


\section{SPS1a' Inspired Fit}
\label{sec:SPS1aPrimeFit}

The SPS1a' inspired scenario studied in this article is defined by the
following high-scale parameters \cite{ref:SPAPaper}:
\begin{eqnarray}
    m_0 & = & 70 \;\mbox{GeV} \hspace{3mm} \mbox{universal scalar mass}
                                                \label{eq:sps1aprimestart}\\
    m_{1/2} & = & 250 \;\mbox{GeV} \hspace{3mm} \mbox{universal gaugino mass}\\
    A_0 & = & -300 \;\mbox{GeV} \hspace{3mm} \mbox{universal trilinear coupling}\\
    \tan \beta & = & 10\\
    \mbox{sign}(\mu) & = & +1 \label{eq:sps1aprimeend}
\end{eqnarray}
As opposed to SPS1a \cite{ref:SPS}, SPS1a' has the advantage that it
is fully consistent with all available measurements including cosmological
data \cite{ref:WMAP}.

Although in the definition of this scenario, gravity mediated SUSY
breaking (mSUGRA) is assumed, no assumption on the SUSY breaking
mechanism is made in the reconstruction of the Lagrangian parameters.
This generality entails the introduction of many soft SUSY breaking
parameters. The advantage of this approach is that the parameters
are reconstructed at the low-energy scale without unnecessary
assumptions and can subsequently be extrapolated to the high-scale to
learn about the SUSY breaking mechanism (``bottom-up'' approach)
\cite{msugrafits}.

\subsection{Fit Assumptions}
\label{sec:assumptions}

The current version of Fittino (version 1.1.1) is able to fit all
low-energy SUSY parameters of theories fulfilling the following
properties:
\begin{itemize}
    \item There is no CP violation in the SUSY sector of the theory,
          i.~e.~all phases vanish.
    \item No inter-generation mixing is present.
    \item Mixing within the first and second generation is zero.
\end{itemize}
With these assumptions, 24 SUSY parameters remain from the initial 105
parameters in the general case.

For the fit presented here, the 24 MSSM parameters available in
Fittino have been further reduced.  We assume that the first and second 
generation sparticle masses are almost degenerate which is motivated
by the fact that no deviation from SM predictions have been found up
to now in low energy data, e.~g.~in kaon physics 
\cite{Eidelman:2004wy,Gabbiani:1996hi}.
 In the squark sector, one unified squark mass parameter
$M_{\tilde{q}_L}$ is assumed for the superpartners of the left-handed
u, d, s and c quarks and one mass parameter $M_{\tilde{q}_R}$ is used
for the superpartners of the right-handed light quarks. Using all
assumtions, 18 free MSSM parameters are left. For the fit, the
low-energy parameters calculated from
equations~\ref{eq:sps1aprimestart} to \ref{eq:sps1aprimeend} are
slightly modified so that this unification is exact. Additionally, the
top quark mass $m_t$ is fitted to account for parametric
uncertainties.

Instead of the trilinear couplings $A_{\text{t}}$,
$A_{\text{b}}$ and $A_{\tau}$, the parameters
\begin{eqnarray}
    X_{\text{t}} & = & A_{\text{t}} - \mu / \tan \beta \label{eq:XATrafo1}\\
    X_{\text{b}} & = & A_{\text{b}} - \mu \tan \beta \label{eq:XATrafo2}\\
    X_{\tau}     & = & A_{\tau} - \mu \tan \beta \label{eq:XATrafo3}
\end{eqnarray}
are fitted in order to reduce correlations with $\tan \beta$ and
$\mu$.

\subsection{Input Observables to the Fit}
\label{sec:Inputs}

A number of anticipated LHC and ILC measurements serve as input
observables to the fit.  For the ILC, running at centre-of-mass
energies of 400 GeV, 500 GeV and 1 TeV is considered with 80~\%
electron and 60~\% positron polarization.  The predicted values of the
observables are calculated using the following prescriptions:
\begin{itemize}
\item {\bf Masses:} \newline The experimental uncertainties of the
  mass measurements are taken from \cite{ref:LHCILC}. For the LHC, the
  mass information is extracted from measurements of edge positions in
  mass spectra. In the gaugino and Higgs sector the precision is
  driven by the ILC, for strongly interacting SUSY particles by the
  LHC. The benefit of combined analyses at LHC and ILC is taken into
  account.
\item {\bf Cross-sections:} \newline Only $\text{e}^+ \text{e}^-$
  cross-sections are included in the fit. However, the measurement of
  absolute cross-sections is impossible for many channels, in which
  only a fraction of the final states can be reconstructed. Therefore,
  absolute cross-section measurements are only used for the
  Higgs-strahlung production of the light Higgs boson, which is
  studied in detail in~\cite{ref:HiggsBr}.
\item {\bf Cross-sections times branching fractions:} \newline Since
  no comprehensive study of the precision of cross-section times
  branching fraction measurements is available, the uncertainty is
  assumed to be the error of a counting experiment with the following
  assumptions:
  \begin{itemize}
  \item The selection efficiency amounts to 50~\%.
  \item 80~\% polarization of the electron beam and 60~\%
    polarization of the positron can be achieved.
  \item 500~fb$^{-1}$ per centre-of-mass energy and polarization
    is collected.
  \item The relative precision is not allowed to be better than 1~\%
    and the absolute accuracy is at most 0.1~fb to account for
    systematic uncertainties.
  \end{itemize}
  All production processes and decays of SUSY particles and Higgs
  bosons are used which have a cross-section times branching ratio
  value of more than 1~fb in one of the $\text{e}^+ \text{e}^-$
  polarization states LL, RR, LR, RL at $500$~GeV
  and LR or RL at $\sqrt{s}=400$~GeV and $\sqrt{s}=1000$~GeV.
\item {\bf Branching fractions:} \newline The four largest branching
  fractions of the lightest Higgs boson are included. The
  uncertainties on the Higgs branching fractions are taken
  from~\cite{ref:HiggsBr}.
\item {\bf Standard Model parameters:} \newline The present
  uncertainties for $m_{\text{W}}$ and $m_{\text{Z}}$ are used as
  conservative estimates. The uncertainty for the top mass
  $m_{\text{t}}$ is assumed to be 50~MeV~(experimental)~\cite{ref:HiggsBr}
  and 100~MeV~(theoretical)~\cite{ref:TopMassTheo}.
\item {\bf Mixing angles:} \newline For tree-level estimates of $\tan
  \beta$, $\mu$ and the parameters of the gaugino sector the chargino
  mixing angles $\cos 2\phi_L$ and $\cos 2\phi_R$ are used. Their
  values are reconstructed using tree-level relations from chargino
  production cross-sections at different beam polarizations
  \cite{Choi:2000ta,Choi:2001ww}.  No use is made of those observables
  in the fit.  The fit result is independent of their assumed
  uncertainties.
\end{itemize}

In order to check the influence of theoretical uncertainties on the
fit results, the fit has been performed twice, once with experimental
uncertainties only and a second time with experimental and estimated
theoretical uncertainties. To be conservative we take present
theoretical uncertainties.  As an estimate, the scale uncertainties as
given in~\cite{ref:SPAPaper} are used for the mass predictions
which also induce a shift in the 'edge' observables. They are
obtained by varying the scale, where the parameters are decoupled from
the RGE running and the shifts to the pole masses are calculated, from
$m_{\text{Z}}$ to 1~TeV. In this way one gets an estimate of the missing
higher order corrections for the shift from the running mass to the
pole mass. Here we use the complete 1-loop formulae given in
\cite{Pierce:1996zz} for the SUSY masses and in addition the 2-loop
corrections for the Higgs sector as given in \cite{Degrassi:2001yf}.
The experimental and theoretical contributions are added in
quadrature. The assumed experimental and theoretical uncertainties for
the masses are listed in Tables~\ref{tab:InputsSPS1aPrime} and
\ref{tab:TheoInputsSPS1aPrime}. For the cross-section and
cross-section times branching fraction measurements, the smallest
allowed relative precision is raised to 2 \% for the fit including
theoretical uncertainties (as opposed to 1 \% for the fit without
theoretical uncertainties). The full list of observables used in the
fit and their uncertainties can be obtained from~\cite{ref:Fittino}.

\section{Fit Results}
\label{sec:UnsmearedFitResultsSPS1aPrime}

The input observables described in Section~\ref{sec:Inputs} are used
to determine the SUSY Lagrangian parameters in a global fit under the
assumptions mentioned in Section~\ref{sec:assumptions}. In total 18
SUSY parameters remain to be fitted. In addition to those, the top
mass $m_{\text{t}}$ is fitted, since it has a relatively large
uncertainty and strongly influences parts of the MSSM observables.
Thus 19 Lagrangian parameters are simultaneously determined in this
fit.

\begin{table}[t]
\begin{center}
\begin{tabular}{l d{7} d{7} d{7} d{7}}
\hline
\multicolumn{1}{c}{Parameter} & \multicolumn{1}{c}{``True'' value} & 
\multicolumn{1}{c}{Fit value} & \multicolumn{1}{r}{Uncertainty} &
\multicolumn{1}{r}{Uncertainty} \\
 & & & \multicolumn{1}{c}{(exp.)} & \multicolumn{1}{c}{(exp.+theor.)}\\
 \hline
$\tan\beta$          &  10.00                &   10.00                & 0.11 & 0.15 \\
$\mu$                &  400.4  \;\text{GeV}  &  400.4   \;\text{GeV}  & 1.2\;\text{GeV} & 1.3\;\text{GeV} \\
$X_{\tau}$           &  -4449.  \;\text{GeV}  & -4449.    \;\text{GeV}  & 20.\;\text{GeV} & 29. \;\text{GeV} \\
$M_{\tilde{e}_R}$    &  115.60  \;\text{GeV} &  115.60 \;\text{GeV}   &  0.13 \;\text{GeV} & 0.43   \;\text{GeV} \\
$M_{\tilde{\tau}_R}$ &  109.89  \;\text{GeV} &  109.89   \;\text{GeV} &  0.32  \;\text{GeV} & 0.56    \;\text{GeV} \\
$M_{\tilde{e}_L}$    &  181.30  \;\text{GeV} &  181.30   \;\text{GeV} &  0.06  \;\text{GeV} & 0.09    \;\text{GeV} \\
$M_{\tilde{\tau}_L}$ &  179.54  \;\text{GeV} &  179.54   \;\text{GeV} &  0.12  \;\text{GeV} & 0.17    \;\text{GeV} \\
$X_{\text{t}}$     &  -565.7  \;\text{GeV} & -565.7    \;\text{GeV} & 6.3   \;\text{GeV} & 15.8   \;\text{GeV} \\
$X_{\text{b}}$  &  -4935. \;\text{GeV} & -4935.   \;\text{GeV} & 1207. \;\text{GeV} & 1713.  \;\text{GeV} \\
$M_{\tilde{q}_R}$    &  503.   \;\text{GeV} &   504.   \;\text{GeV} & 12.  \;\text{GeV} &  16.  \;\text{GeV} \\
$M_{\tilde{b}_R}$    &  497.   \;\text{GeV} &  497.    \;\text{GeV} & 8.  \;\text{GeV} &  16.  \;\text{GeV} \\
$M_{\tilde{t}_R}$    &  380.9   \;\text{GeV} &  380.9    \;\text{GeV} & 2.5  \;\text{GeV} &  3.7    \;\text{GeV} \\
$M_{\tilde{q}_L}$    &  523.   \;\text{GeV} &  523.    \;\text{GeV} &  3.2 \;\text{GeV} & 4.3    \;\text{GeV} \\
$M_{\tilde{t}_L}$    &  467.7   \;\text{GeV} &  467.7    \;\text{GeV} & 3.1 \;\text{GeV} & 5.1    \;\text{GeV} \\
$M_1$                &  103.27 \;\text{GeV} &  103.27  \;\text{GeV} & 0.06  \;\text{GeV} & 0.14   \;\text{GeV} \\
$M_2$                &  193.45 \;\text{GeV} &  193.45  \;\text{GeV} & 0.08  \;\text{GeV} & 0.13  \;\text{GeV} \\
$M_3$                &  569.  \;\text{GeV} &  569.    \;\text{GeV} &  7.  \;\text{GeV} &   7.4   \;\text{GeV} \\
$m_{\text{A}_{\text{run}}}$ & 312.0 \;\text{GeV} & 311.9 \;\text{GeV} & 4.3  \;\text{GeV} & 6.5   \;\text{GeV} \\
$m_{\text{t}}$     &  178.00  \;\text{GeV} &  178.00   \;\text{GeV} &  0.05  \;\text{GeV} & 0.12  \;\text{GeV} \\
\hline
\multicolumn{5}{l}{Corresponding values for the trilinear couplings:}\\
\hline
$A_{\tau}$         & -445.\;\text{GeV} & -445.\;\text{GeV} & 40. \;\text{GeV} & 52.\;\text{GeV} \\
$A_{\text{t}}$         & -526.\;\text{GeV} & -526.\;\text{GeV} & 6.\;\text{GeV} & 16.\;\text{GeV} \\
$A_{\text{b}}$         & -931. \;\text{GeV} & -931.\;\text{GeV} & 1184. \;\text{GeV} & 1676.\;\text{GeV} \\
\hline
\multicolumn{5}{c}{$\chi^2$ for unsmeared observables: $2.1\times 10^{-5}$}\\
\hline
\end{tabular}
\end{center}
\caption{The Fittino fit result for the SPS1a' inspired scenario.
  The left column shows the values predicted by SPheno version 2.2.2
  for this scenario, the second column exhibits the central values of
  the fit with unsmeared input observables, the third column displays
  the parameter uncertainties for the fit with experimental
  uncertainties only.  The last column shows the parameter
  uncertainties for the fit with experimental and theoretical
  uncertainties. In addition to the fitted parameters, the corresponding
  values for the trilinear couplings are given as obtained from reversing
  the parameter transformations~(\ref{eq:XATrafo1})-(\ref{eq:XATrafo3}),
  taking correlations into account.\vspace{2mm}}
\label{tab:UnsmearedFitResultSPS1aPrime}
\end{table}

\begin{figure}[t]
    \begin{center}
      \includegraphics[width=0.9\textwidth]{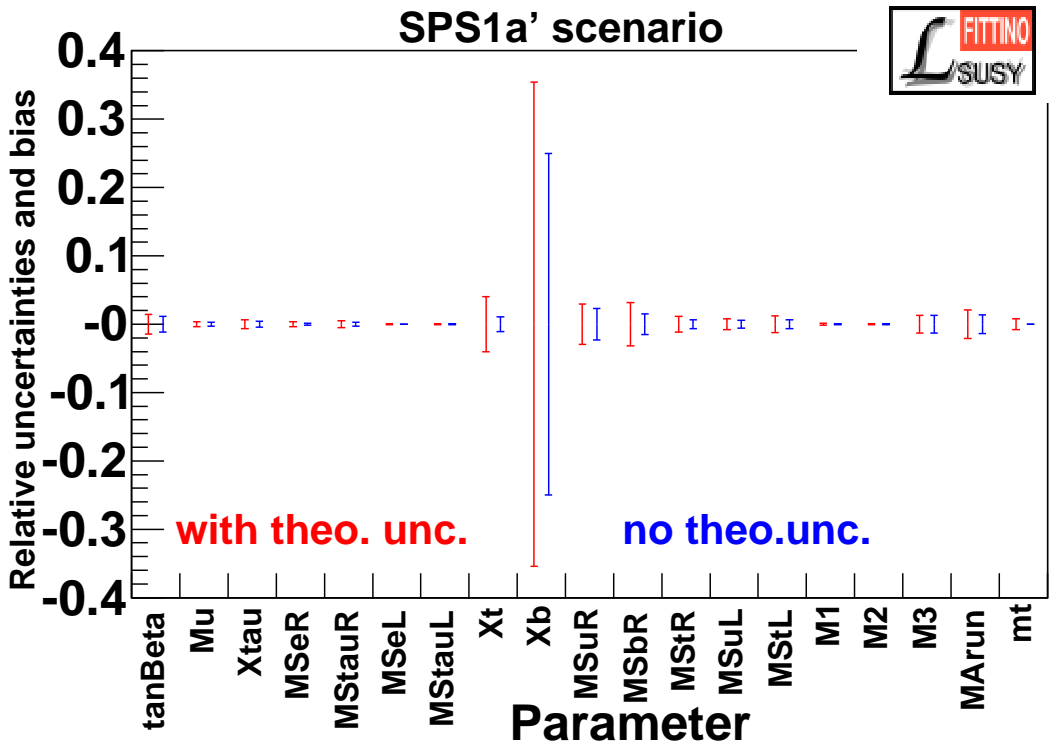}
    \end{center}
\caption{Relative uncertainties of the parameter measurements with
	(red, left error bars) and without (blue, right error bars)
	theoretical uncertainties.}
\label{fig:RelativeUncertainties}
\end{figure}

As shown in Table~\ref{tab:UnsmearedFitResultSPS1aPrime} all
parameters are perfectly reconstructed at their input values. Due to
the fact that the input observables are unsmeared in this fit, the
final $\chi^2$ is close to zero at $\chi^2 = 2.1 \times 10^{-5}$.

\subsection{Extraction of the Fit Uncertainties and Correlations}

After a successful convergence of the simulated annealing algorithm to
the input parameter values, the fit uncertainties are evaluated by
carrying out many individual fits with input values randomly smeared
within their uncertainty range using a Gaussian probability
density. The complete covariance matrix and the correlation matrix are
derived from the spread of the fitted parameter values.

For the case without and for the case with theoretical uncertainties
about 1000 fits are performed. For each of those fits, the $\chi^2$
is minimized. For a large and complex parameter space with large
correlations among the parameters this method has turned out to be
more robust than a MINOS error analysis.

\begin{figure}[t]
    \begin{center}
        \includegraphics[width=0.48\textwidth]{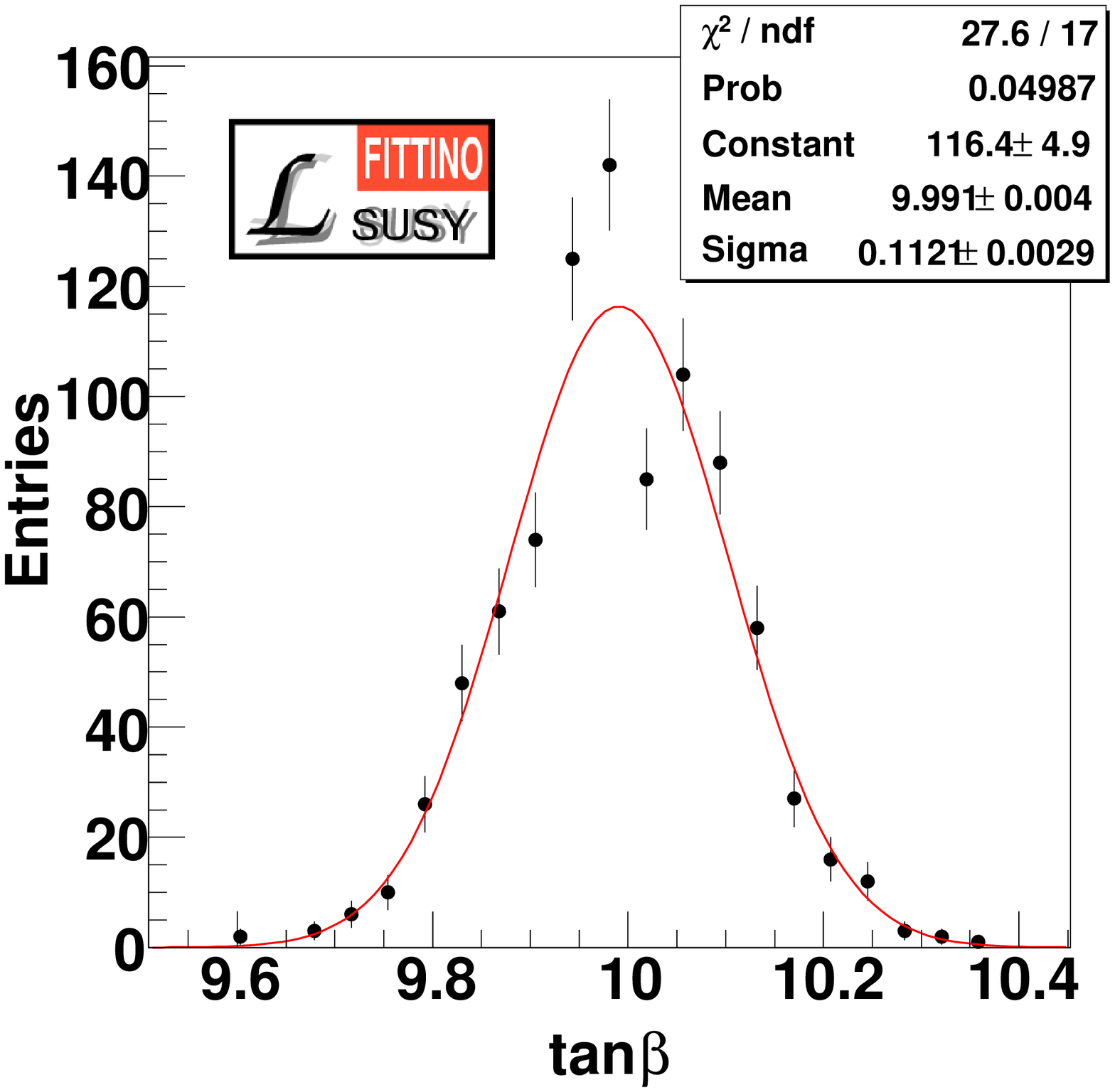}
        \hfill              
        \includegraphics[width=0.48\textwidth]{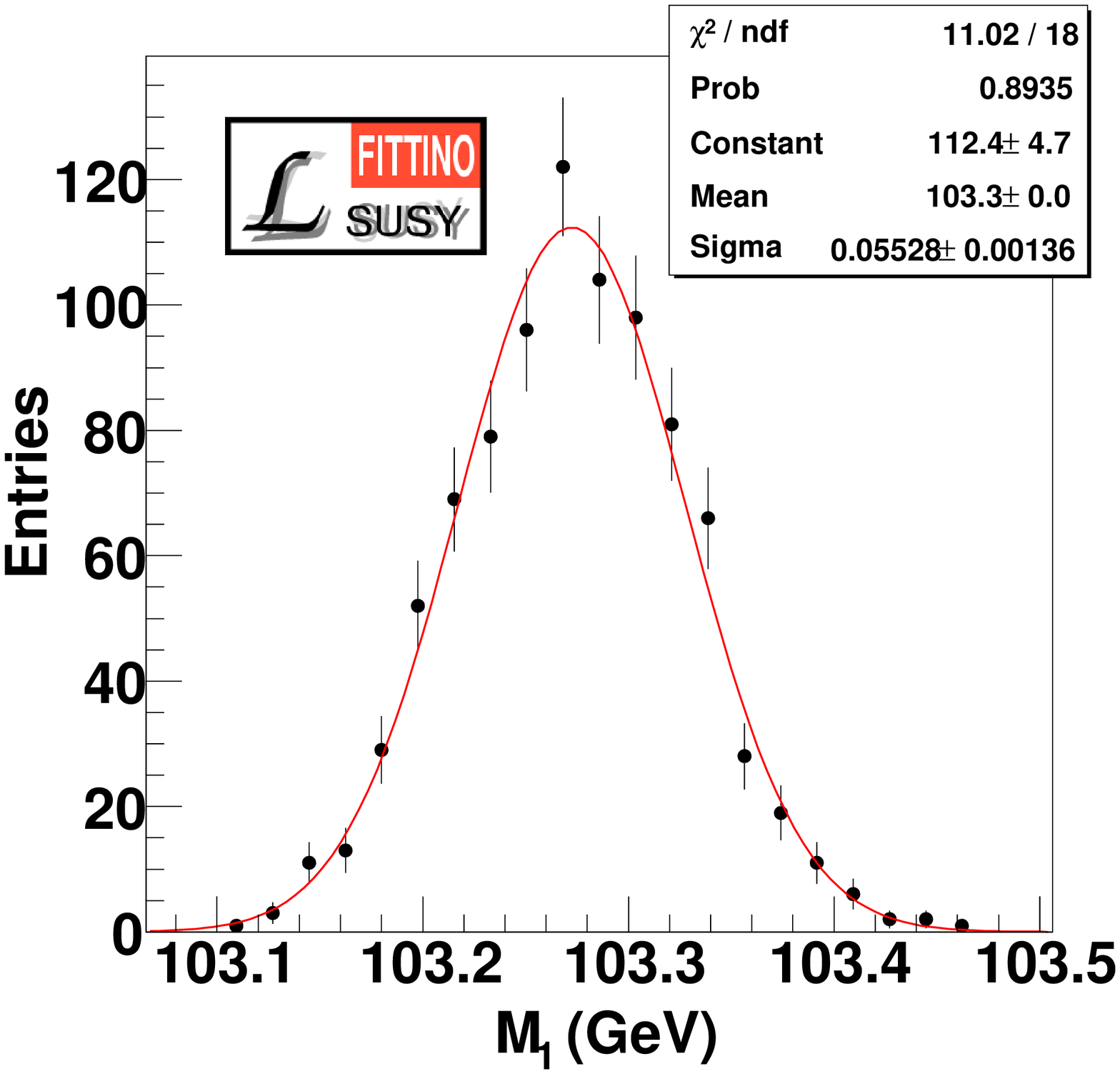}\\
        \includegraphics[width=0.48\textwidth]{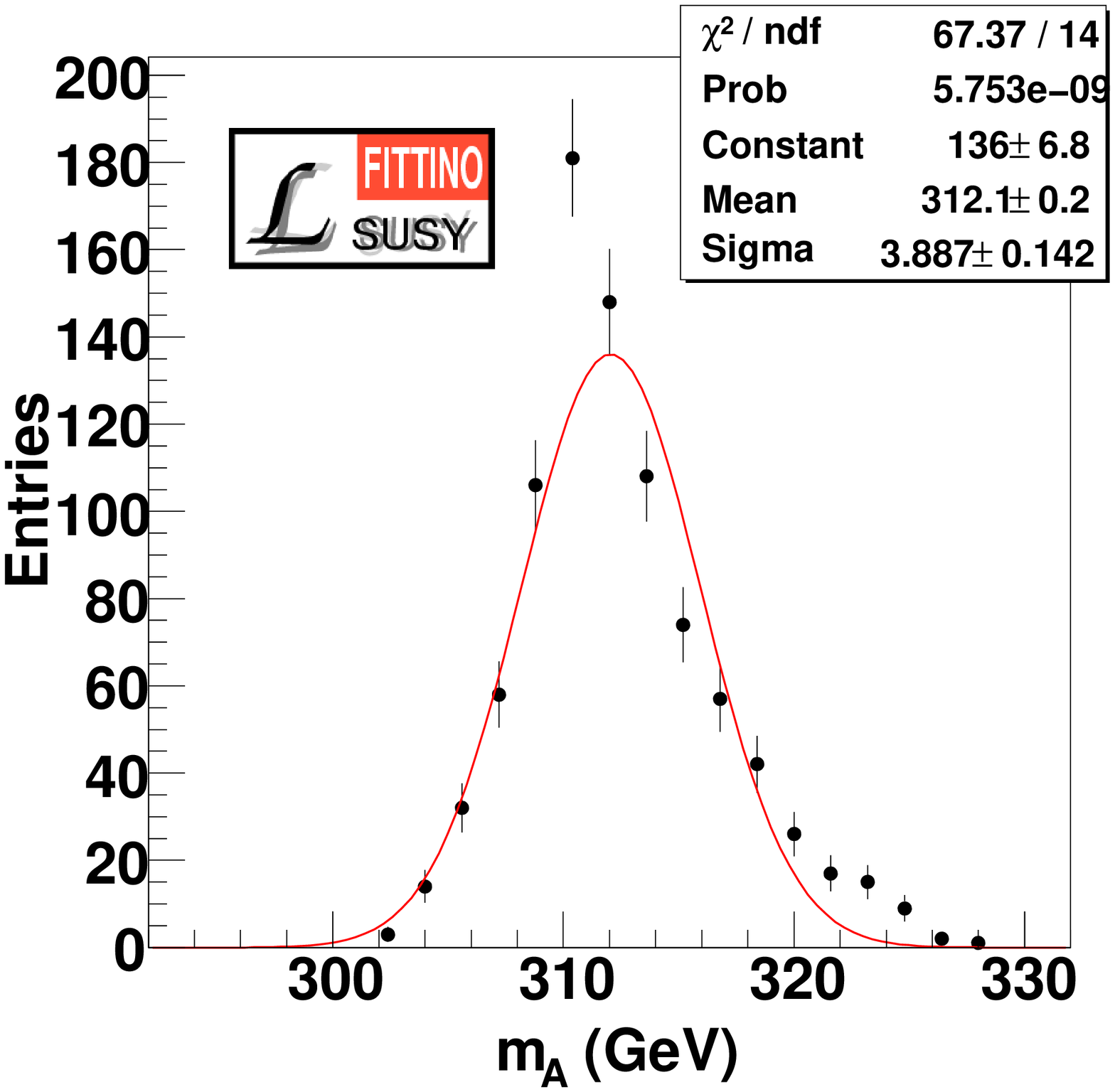}
        \hfill              
        \includegraphics[width=0.48\textwidth]{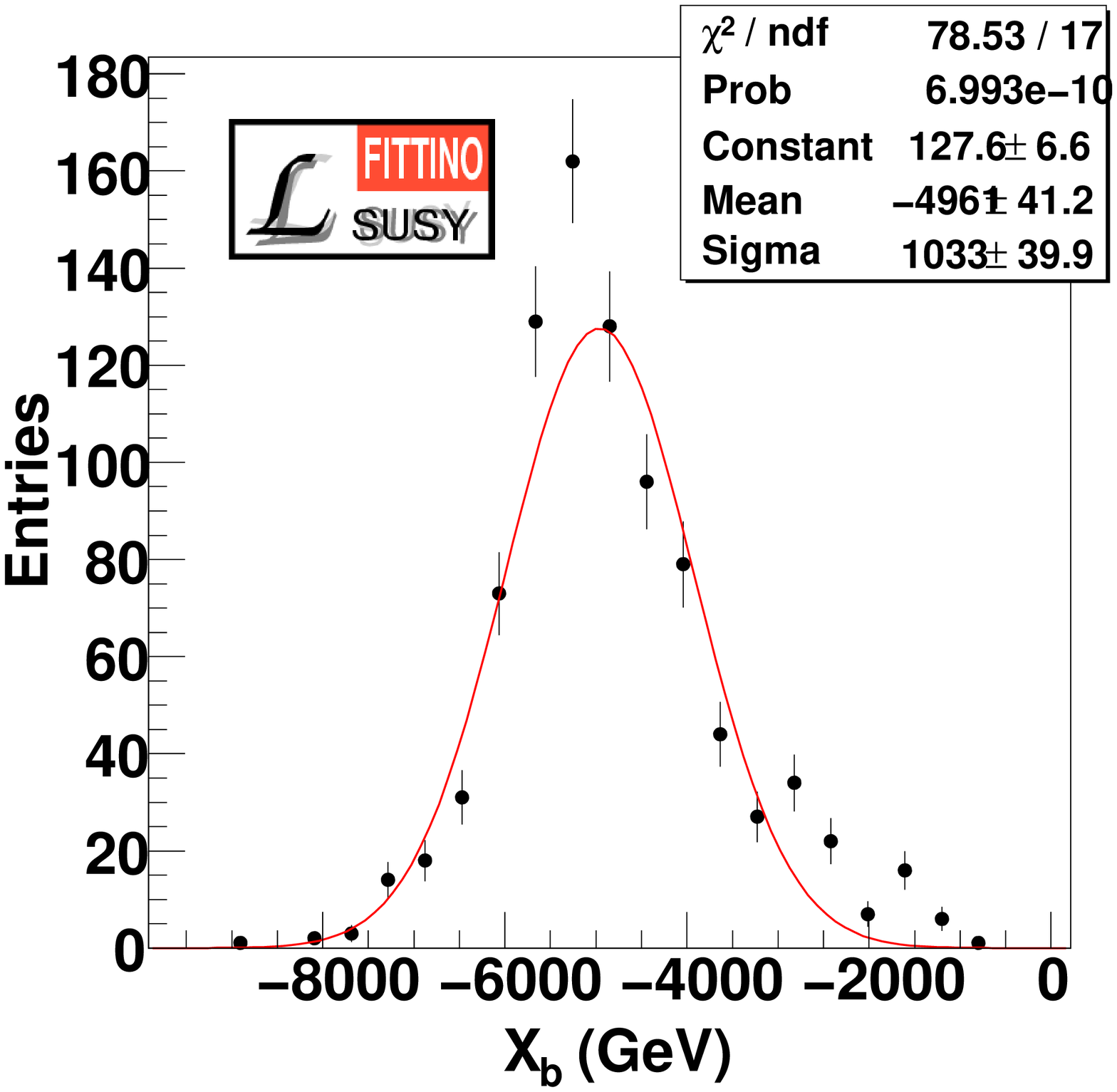}
    \end{center}
\caption{Examples of the toy fit value distributions for $\sim$1000 independent fits
  with observables smeared within their uncertainties.}
\label{fig:pulls4}
\end{figure}

An example of the outcome of this procedure for the fit (with
experimental uncertainties only) from
Section~\ref{sec:UnsmearedFitResultsSPS1aPrime} is shown in
Figure~\ref{fig:pulls4} for the parameters $\tan\beta$, $M_1$,
$m_{\mathrm{A}_{\mathrm{pole}}}$ and $X_{\text{b}}$ for the $\sim$1000
independent fits.  All distributions apart from
$m_{\mathrm{A}_{\mathrm{pole}}}$ and $X_{\text{b}}$ agree well with
Gaussians.

The difference of the RMS of the parameter distribution and the width
of a Gaussian fitted to the distribution (normalized to the Gaussian
width) is shown in Tab.~\ref{tab:GoodnessOfFit} for the 19 parameters.
The distributions showing the largest disagreement are
$m_{\mathrm{A}_{\mathrm{pole}}}$ and $X_{\text{b}}$, indicating that a
parabolic error assumption for these parameters is not completely
correct.


The parameter uncertainties extracted from the fit value distributions
are shown in Tab.~\ref{tab:UnsmearedFitResultSPS1aPrime}, while the
corresponding correlation matrix is displayed in
Tab.~\ref{tab:CorrelationMatrix1} and
\ref{tab:CorrelationMatrix2}. Additionally, a graphical representation
of the relative uncertainties of the individual parameters, both with
and without theoretical uncertainties, is given in
Figure~\ref{fig:RelativeUncertainties}.

\begin{table}[t]
  \begin{center}
    \begin{tabular}{l d{7} l d{7}}
      \hline
      Parameter & \multicolumn{1}{c}{$(\text{RMS}-\sigma)/\sigma$} & Parameter & 
      \multicolumn{1}{c}{$(\text{RMS}-\sigma)/\sigma$} \\
      \hline
      $\tan\beta$                 & 0.01 & $M_{\tilde{b}_R}$           &  0.07 \\
      $\mu$                       & 0.06 & $M_{\tilde{t}_R}$           &  0.02 \\
      $X_{\tau}$                  & 0.03 & $M_{\tilde{q}_L}$           &  0.01 \\
      $M_{\tilde{e}_R}$           & 0.02 & $M_{\tilde{t}_L}$           &  0.04 \\
      $M_{\tilde{\tau}_R}$        & 0.00 & $M_1$                       &  0.01 \\
      $M_{\tilde{e}_L}$           & 0.02 & $M_2$                       & -0.01 \\
      $M_{\tilde{\tau}_L}$        & 0.01 & $M_3$                       &  0.04 \\
      $X_{\text{t}}$              & 0.03 & $m_{\text{A}_{\text{run}}}$ &  0.14 \\
      $X_{\text{b}}$              & 0.17 & $m_{\text{t}}$              &  0.01 \\
      $M_{\tilde{q}_R}$           & 0.04 & \\
      \hline
    \end{tabular}
  \end{center}
  \caption{Comparison of the RMS and the Gaussian width $\sigma$ fitted to the parameter
  distributions.\vspace{2mm}}
  \label{tab:GoodnessOfFit}
\end{table}

\afterpage{\clearpage}

\begin{figure}[t]
    \begin{center}
        \includegraphics[width=0.7\textwidth]{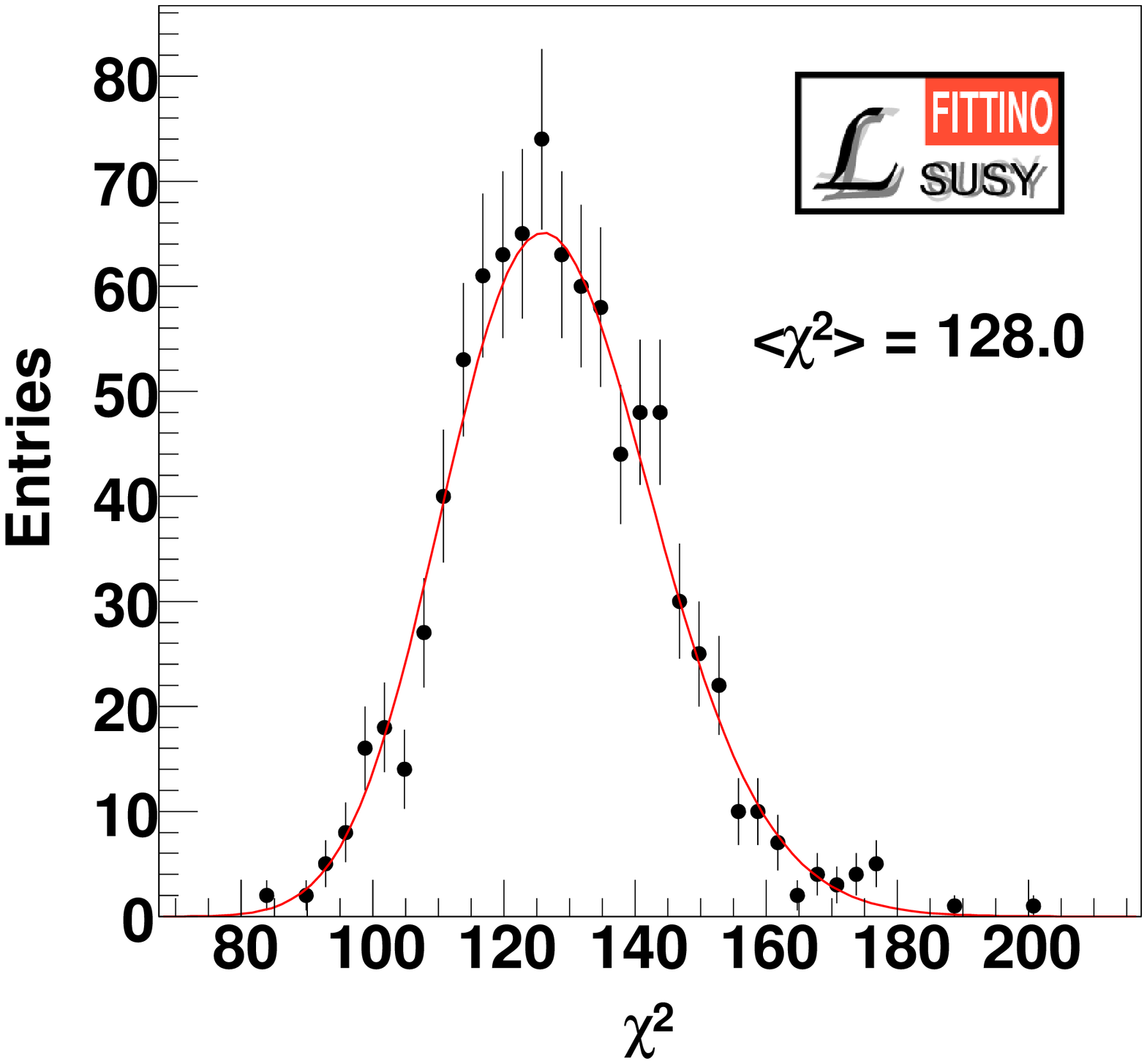}
    \end{center}
\caption{$\chi^2$ distribution for $\sim$1000 independent fits with
  observables smeared within their uncertainties. The mean $\chi^2$
  from a $\chi^2$ function fitted to the observed distribution of
  128.0 agrees with the expectation of 129.0 $\pm$ 0.7.}
\label{fig:chi2}
\end{figure}

Figure~\ref{fig:chi2} shows the distribution of the $\chi^2$ values
for these $\sim$1000 independent fits for the case without theoretical
uncertainties.  The mean $\chi^2$ obtained from a fit of the $\chi^2$
distribution to the observed distribution is 128.0, in agreement with
the expectation of 129.0~$\pm$~0.7.  This shows that the fits
converge well to the true minimum of the $\chi^2$ surface for the
smeared observables, implying that the uncertainties extracted from
the toy fit value distributions are correct.

\subsection{Interpretation of the Fit Results}

Most parameters are reconstructed to a precision better than or around
1~\%. For the U(1) gaugino mass parameter $M_1$, an accuracy below the
per mil level is achieved for the fit including experimental
uncertainties only.  But its precision is worse by more than a factor
of 2 once theoretical uncertainties are included. The uncertainties on
other parameters increase by up to a factor of 3.3 if theoretical
uncertainties are taken into account, such as $M_{\tilde{\text{e}}_R}$. In order
to fully benefit from the precision data provided in particular by the
ILC, more precise calculations would be beneficial.

The least constrained variables are the mixing parameters. Especially
$X_{\text{b}}$ has only small influence on the Higgs sector and the
sbottom masses, resulting in a relative precision of only 25~\%
(corresponding to 130~\% relative uncertainty for $A_{\text{b}}$). The
situation is better for $X_{\tau}$ and $X_{\text{t}}$. The high
precision in the chargino and slepton sector as well as the occurrence
of the decay channel $\text{H}^+ \rightarrow \tilde{\nu}_{\tau_L}
\tilde{\tau}_1^+$ allows to determine $X_{\tau}$ to an accurary of
less than 0.5~\% (corresponds to 9~\% for $A_{\tau}$).  The achieved
relative precision of 1~\% for $X_{\text{t}}$ (1~\% for $A_{\text{t}}$)
is possible due to the large mass splitting in the stop sector and its
influence on the Higgs sector.  Searching for sensitive observables to
constrain the mixing parameters to a larger extent is important to
improve their precisions.

The benefit of a global fit is that the full correlation matrix of all
parameters is available (see
Appendix~\ref{sec:CorrelationMatrixAppendix}).  As expected
correlations within sectors can be strong. The parameters of the
gaugino sector $\tan \beta$, $\mu$, $M_1$ and $M_2$, for example,
reveal correlations up to 69 \%. But even between different sectors
the correlations cannot be neglected. One example is
$M_{\tilde{\text{q}}_R}$ from the squark sector whose correlation
coefficient with $M_{\tilde{\text{e}}_R}$ from slepton sector amounts
to $-0.79$. Even the relatively robust gaugino sector is affected by
such inter-sector correlations. $M_2$, for example, has a correlation
of 20~\% with $M_{\tilde{\text{q}}_L}$ from the sfermion sector. This
underlines the importance of fitting all SUSY parameters
simultaneously.  Neglecting these inter-sector correlations causes
incorrect fit uncertainties in fits restricted to specific sectors.

\subsection{Fits in Subspaces}

This section is devoted to a check of the influence of restricted fits
on the precision of the reconstructed parameters and their
uncertainties. It might be interesting to study the possibility to
extract limited information on the SUSY parameters in a certain
subspace only, because the input described in Section~\ref{sec:Inputs}
will not be completely available at the beginning of ILC running at
lower energies. Therefore fits with inputs restricted to the
low-energy running phase of ILC have been performed. But this entails
fixing parameters which cannot be determined with the low-energy
data. The following excerpt of observables from
Section~\ref{sec:Inputs} has been used for the fit:
\begin{itemize}
  \item All SM observables.
  \item All mass measurements in the gaugino and slepton sector from
    ILC running at $\sqrt{s} = 500$ GeV.
  \item All chargino, neutralino and $\tilde{e}$ cross-section times
    branching fraction measurements at $\sqrt{s} = 400$ GeV and
    $\sqrt{s} = 500$ GeV.
\end{itemize}
No Higgs sector parameters and observables are used in the fit, since
they have large dependencies on squark sector parameters such as
$X_{\text{t}}$ and $M_{\tilde{t}}$.

The fixed and fitted parameters are listed in
Table~\ref{tab:PartialFit}.  All fixed parameters are set to the
correct values in a first fit.  As shown in Table~\ref{tab:PartialFit}
the central values of the fit reproduce the correct parameter
values. In a second step the fit is performed with all fixed
parameters set to modified values. The chosen numbers are estimates in
a situation where the strongly interacting sector has not been
precisely measured yet. The quality of the fit results for this
second fit depends on the chosen set of fixed and fitted parameters.

\begin{table}[t]
\begin{center}
\begin{tabular}{l d{-1} d{-1} d{-1} d{7}}
\hline
Parameter & \multicolumn{1}{c}{``True'' value} & \multicolumn{1}{c}{Fit with correctly} 
& \multicolumn{1}{c}{Fit with incorrectly} & \multicolumn{1}{c}{Uncertainty} \\
& & \multicolumn{1}{c}{fixed parameters} & \multicolumn{1}{c}{fixed parameters}
& \\
\hline
\multicolumn{5}{c}{Fixed parameters}\\
\hline
$X_{\text{t}}$              &  -565.7  \;\text{GeV} &  -565.7  \;\text{GeV} &     -30.0  \;\text{GeV} & \multicolumn{1}{c}{fixed} \\
$X_{\text{b}}$              & -4934.8  \;\text{GeV} & -4934.8  \;\text{GeV} &   -4000.0  \;\text{GeV} & \multicolumn{1}{c}{fixed} \\
$M_{\tilde{q}_R}$           &   501.6  \;\text{GeV} &   501.6  \;\text{GeV} &     600.0  \;\text{GeV} & \multicolumn{1}{c}{fixed} \\
$M_{\tilde{b}_R}$           &   497.4  \;\text{GeV} &   497.4  \;\text{GeV} &     600.0  \;\text{GeV} & \multicolumn{1}{c}{fixed} \\
$M_{\tilde{t}_R}$           &   503.9  \;\text{GeV} &   503.9  \;\text{GeV} &     500.0  \;\text{GeV} & \multicolumn{1}{c}{fixed} \\
$M_{\tilde{q}_L}$           &   523.2  \;\text{GeV} &   523.2  \;\text{GeV} &     500.0  \;\text{GeV} & \multicolumn{1}{c}{fixed} \\
$M_{\tilde{t}_L}$           &   467.7  \;\text{GeV} &   467.7  \;\text{GeV} &     500.0  \;\text{GeV} & \multicolumn{1}{c}{fixed} \\
$M_3$                       &   568.9  \;\text{GeV} &   568.9  \;\text{GeV} &     700.0  \;\text{GeV} & \multicolumn{1}{c}{fixed} \\
$m_{\text{A}_{\text{run}}}$ &   312.0  \;\text{GeV} &   312.0  \;\text{GeV} &     400.0  \;\text{GeV} & \multicolumn{1}{c}{fixed} \\
$m_{\text{t}}$              &   178.0  \;\text{GeV} &   178.0  \;\text{GeV} &     178.0  \;\text{GeV} & \multicolumn{1}{c}{fixed} \\
\hline
\multicolumn{5}{c}{Fitted parameters}\\
\hline
$\tan \beta$                 &  10.00               &  10.00                &  11.1                &     0.47               \\
$\mu$                        & 400.39  \;\text{GeV} &  400.388 \;\text{GeV} &  388.3  \;\text{GeV} &     3.1   \;\text{GeV} \\
$X_{\tau}$                   & -4449.2 \;\text{GeV} & -4449.2  \;\text{GeV} & -4447.8 \;\text{GeV} &    37.2   \;\text{GeV} \\
$M_{\tilde{e}_R}$            & 115.60  \;\text{GeV} &  115.602 \;\text{GeV} &  113.74 \;\text{GeV} &     0.06  \;\text{GeV} \\
$M_{\tilde{\tau}_R}$         & 109.89  \;\text{GeV} &   109.89 \;\text{GeV} &  107.77 \;\text{GeV} &     0.48  \;\text{GeV} \\
$M_{\tilde{e}_L}$            & 181.30  \;\text{GeV} &  181.304 \;\text{GeV} &  181.76 \;\text{GeV} &     0.04  \;\text{GeV} \\
$M_{\tilde{\tau}_L}$         & 179.54  \;\text{GeV} &   179.54 \;\text{GeV} &  179.99 \;\text{GeV} &     0.14  \;\text{GeV} \\
$M_1$                        & 103.271 \;\text{GeV} &  103.271 \;\text{GeV} &  103.11 \;\text{GeV} &     0.05  \;\text{GeV} \\
$M_2$                        & 193.446 \;\text{GeV} &  193.445 \;\text{GeV} &  193.49 \;\text{GeV} &     0.12  \;\text{GeV} \\
\hline
$\chi^2$               &                      & \multicolumn{1}{c}{$1.8 \times 10^{-5}$} & \multicolumn{1}{c}{$5.89$} & \\
\hline
\end{tabular}
\end{center}
\caption{Fittino fits of a parameter subspace using inputs from the low-energy running of ILC only.\vspace{2mm}}
\label{tab:PartialFit}
\end{table}

This test shows that subspace fits can provide results of the correct
order of magnitude.
Subspace fits are justified at the beginning
of the exploration of SUSY phenomena where experimental uncertainties
are large and only limited data is available for SUSY parameter
analysis. Depending on the chosen set of fixed and fitted
parameters, subspace fits might yield parameter values whose biases
are several times as large as their uncertainties obtained from the
fit.  This unfortunate feature is a consequence of neglected
correlations.  The errors obtained from subsector fits do not express
a probability to find the true parameter values within the uncertainty
range. Uncovering these shifts by looking at the $\chi^2$ of the fit
is unlikely in reality since a degradation of $\Delta \chi^2 = 5.89$,
which is the largest $\chi^2$ enhancement observed in above fits,
might still yield a reasonable $\chi^2$ value for the fit with its 85
degrees of freedom. In any case, as this example shows, a precision
determination of the SUSY Lagrangian parameters requires a
comprehensive global fit using inputs from all sectors of the theory
and fitting parameters from all sectors of the theory.

\afterpage{\clearpage}

\section{Extrapolation to High Energy Scales}
\label{sec:extrapolation}

As soon as the complete set of parameters is known at the electroweak
scale, they can be extrapolated to high energy scales to test
ideas concerning SUSY breaking or grand unification within the
bottom-up approach \cite{msugrafits}. In this way one exploits the
experimental information to the maximum extent possible without any
assumptions on the structure of the high scale theory except that
there are no new particles between the electroweak scale and the high
scale\footnote{In the case that additional heavy particles,  e.~g.~right handed
neutrinos, one is even able to get information on their mass scale in
this approach as has been shown in \cite{Freitas:2005et}.}.

\begin{figure}[th]
\setlength{\unitlength}{1mm}
\begin{picture}(150,65)
\put(0,-70){\epsfig{figure=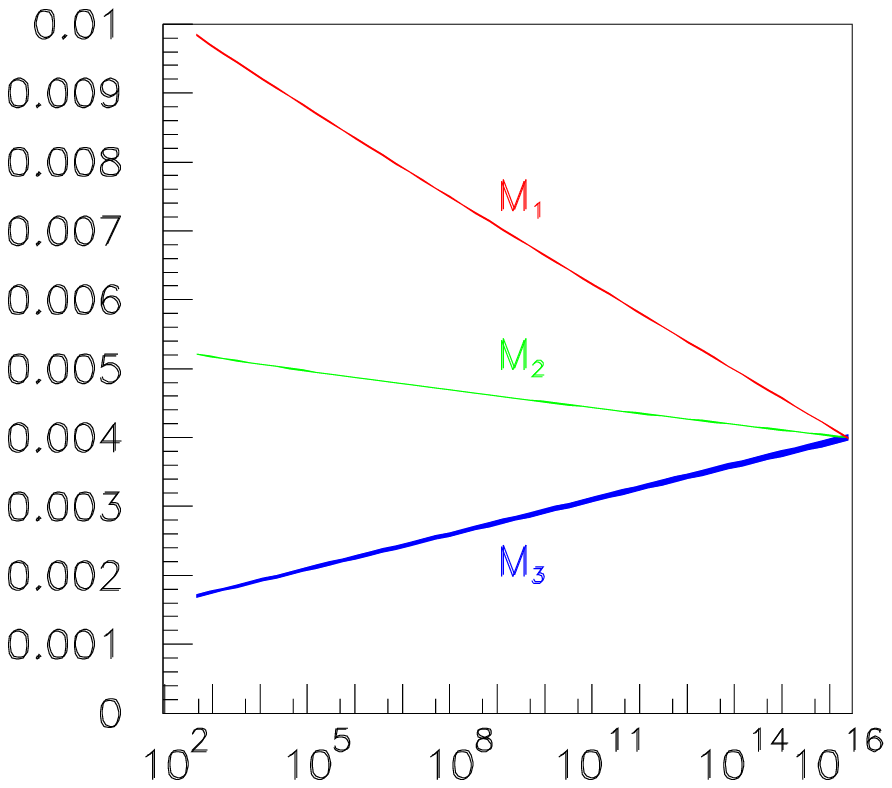,height=140mm,width=110mm}}
\put(0,60){\mbox{ {\bf a)} \small $M^{-1}_i$ [GeV$^{-1}$]}}
\put(35,-2){\mbox{ \small $Q$ [GeV]}}
\put(55,-70){\epsfig{figure=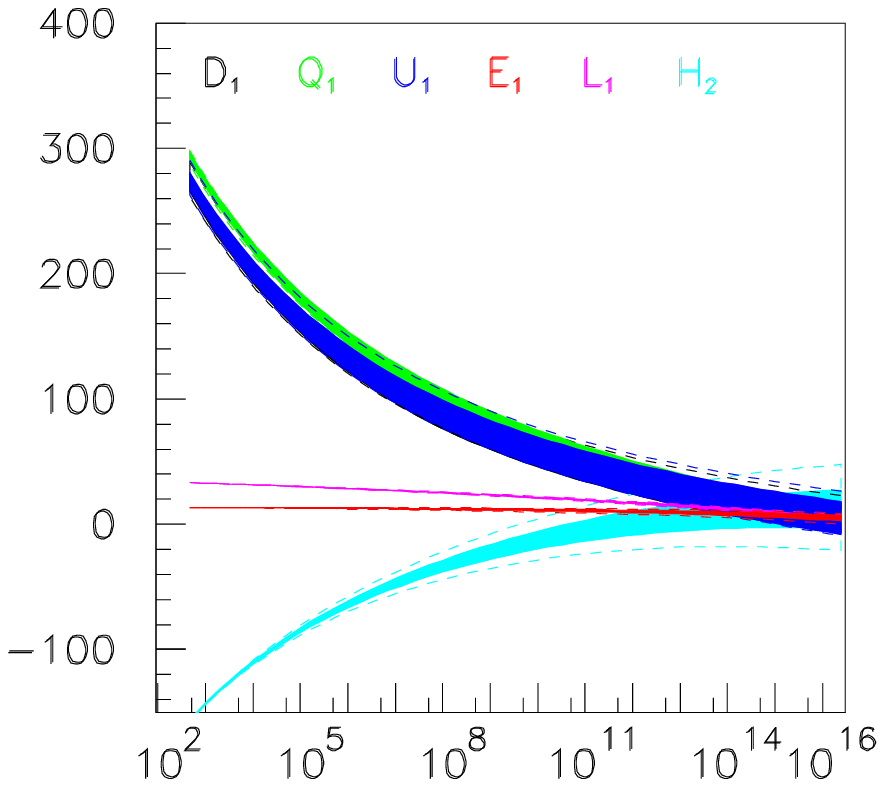,height=140mm,width=110mm}}
\put(55,60){\mbox{ {\bf b)}  \small $M^2_{\tilde j}$ [GeV$^2$]}}
\put(110,-70){\epsfig{figure=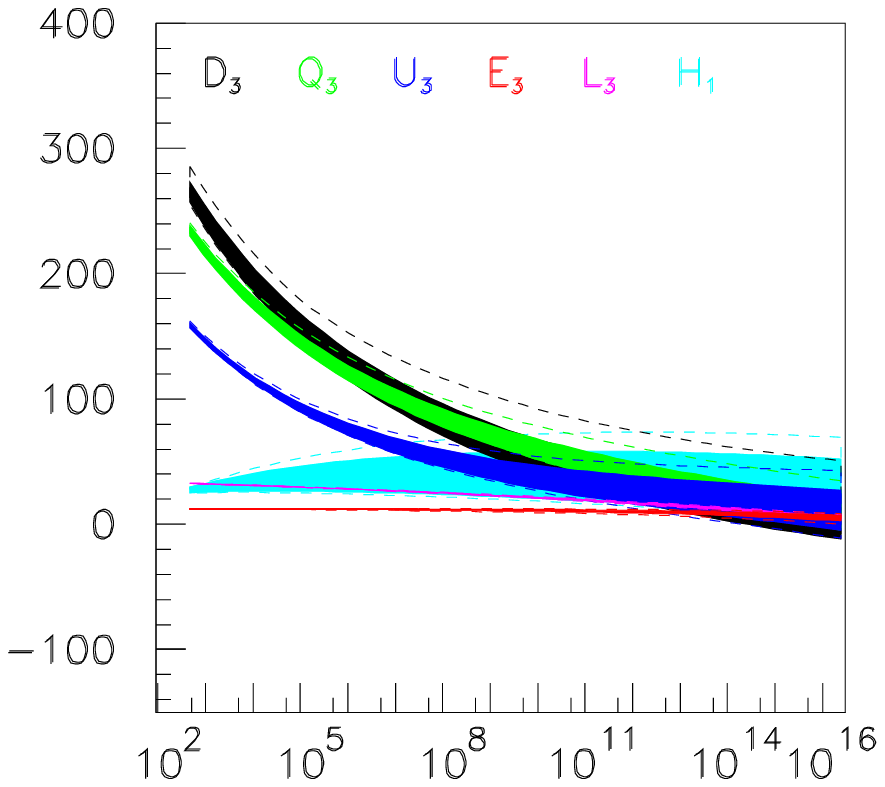,height=140mm,width=110mm}}
\put(90,-2){\mbox{ \small $Q$ [GeV]}}
\put(110,60){\mbox{ {\bf c)} \small $M^2_{\tilde j}$ [GeV$^2$]}}
\put(145,-2){\mbox{ \small $Q$ [GeV]}}
\end{picture}
\caption{
Running of {\bf a)} the gaugino mass parameters,
{\bf  b)}  first generation scalar mass parameters  and $M^2_{H_2}$
and {\bf  c)}  third generation scalar mass parameters  and $M^2_{H_1}$
in SPS1a$'$. 
Full bands: only experimental errors are taken into account;
dashed lines: today's theoretical errors are taken into account
as a conservative estimate.}
\label{fig:running}
\end{figure}

 The parameters at the high energy are related to the electroweak scale
parameters by renormalization group equations (RGEs). Here we use two-loop
RGEs for the evolution \cite{Martin:1993zk}. The evolution of the parameters
from the electroweak scale to the GUT scale are shown in Fig.~\ref{fig:running}
where the GUT scale is defined as the scale where the $SU(2)$ gauge coupling
$g_2$ coincides with $U(1)$ gauge coupling $g_1$ (using the proper normalisation
$g_1 = \sqrt{5/3} g'$) and the value for the electroweak scale is given
by $\sqrt{m_{\tilde t_1}m_{\tilde t_2} }$. The latter value is motivated
by the expectation that at this scale the effect of missing higher order corrections
 in the calculation of $m_{\text{h}}$, which is most probably the most precisely measured
mass, is minimized.

In  Fig.~\ref{fig:running} a the evolution for the gaugino parameters $M^{-1}_i$
is presented. They are clearly under excellent control allowing for
precise tests concerning their unification at $M_{GUT}$. Note, that
their 'unification point' is in principle independent of the one for the
gauge couplings allowing for a cross check if both sets of parameters
indeed meet at the same point.

\begin{figure}[t]
\setlength{\unitlength}{1mm}
\begin{picture}(150,65)
\put(55,-70){\epsfig{figure=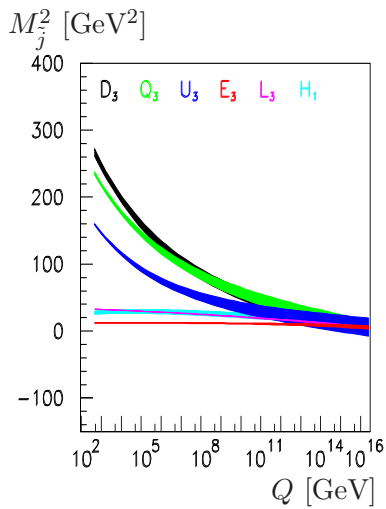,height=140mm,width=110mm}}
\put(55,60){\mbox{ \small $M^2_{\tilde j}$ [GeV$^2$]}}
\put(90,-2){\mbox{ \small $Q$ [GeV]}}
\end{picture}
\caption{
Running of   third generation scalar mass parameters  and $M^2_{H_1}$
in SPS1a$'$ assuming that $A_{\text{b}}$ is known with an 'accuracy' of 50\%.}
\label{fig:running2}
\end{figure}

 In  Fig.~\ref{fig:running} b and c the running of the scalar mass parameters
are shown. The slepton mass parameters are under excellent control whereas
the squark and the Higgs mass parameters are signifcantly worse. Part
of this difference can be traced back to the structure of the RGEs as
explained in detail in the second paper of \cite{msugrafits}. 
In addition the analysis used here does not coincide with ones 
presented in \cite{msugrafits} and \cite{Allanach:2004ud}: 
In \cite{msugrafits} and \cite{Allanach:2004ud} it has been assumed that
all $A$ parameters meet within 1-sigma as up to now no good
proposal exists for a precise measurement of $A_b$. As we did
not use this assumption in this study we find significantly larger errors
for the third generation squark mass parameters and the Higgs parameters.
Using this assumption or, equivalently, having tools at hand which allow
for a better determination of  $A_{\text{b}}$, the error bands
get significantly reduced as can be seen in Fig.~\ref{fig:running2} where
we show the running of the same parameters as in Fig.~\ref{fig:running} c
assuming that $A_{\text{b}}$ is known within 50~\% at the electroweak scale.

In Fig.~\ref{fig:running} we also show the case that today's theoretical
uncertainties are taken into account (dashed lines). In case of the
gaugino mass parameters the effect is so small that it cannot be seen in the plot
(see also Tab.~\ref{tab:UnsmearedFitResultSPS1aPrime}). Note that the
smallness of the theoretical uncertainty on the gluino mass is due
to the fact that the squarks have a similar mass. It gets
larger if there is a significant hierarchy between these particles, e.~g.~in
focus point scenarios. One sees that the main effect is on the parameters
of the third generation sfermions and the Higgs mass parameters. The main reason
for this is that the error on $X_{\text{b}}$ and $M_{\tilde{\text{b}}_R}$ get
considerably larger once the theoretical uncertainties are taken into account.
This figure clearly shows that there is a strong need to further increase
accuracies of the theoretical predictions.

\section{Fit Within mSUGRA Framework}\label{sec:mSUGRAFit}

Instead of determining the full spectrum of the general low-energy
Lagrangian parameters and inferring the high-scale mechanisms from
them, the parameters of a more restrictive high-scale scenario can be
fitted directly to the observables. This section presents an example
for the mSUGRA scenario SPS1a', using the same set of observables as
described in Section~\ref{sec:Inputs}. The parameters of
the scenario are $\tan \beta$, $M_0$, $M_{1/2}$, $A_0$ and
$\text{sign} (\mu)$. The latter has been fixed to 1 for this exercise.

Since there are only four continuous parameters, the fit converges
without complex fitting techniques and the uncertainty determination
is stable.  The fit results for the simulated LHC+ILC measurements
from Section~\ref{sec:UnsmearedFitResultsSPS1aPrime} and observables
set to the exact prediction of the mSUGRA model are given in
Table~\ref{tab:mSUGRAFitResults}, along with results for a fit just
using the LHC observables in the fit. As an additional LHC
measurement, the ratio of Higgs couplings as simulated in
\cite{ref:Duehrssen} has been used. For the LHC$+$ILC measurements,
due to the much more constrained theoretical parameter space, the
precision on the parameters is much better than for the general
low-energy model (see Table~\ref{tab:UnsmearedFitResultSPS1aPrime}).  For
the same reason, the LHC alone is able to perform well constrained
parameter measurements with uncertainties typically one to two orders
of magnitude larger than in the case of the combination of both.

\begin{table}
\begin{center}
\begin{tabular}{l d{7} d{7} d{7} d{7} d{7}}
\hline
& \multicolumn{1}{l}{SPS1a' value} & \multicolumn{1}{l}{Fitted value}
& \multicolumn{1}{l}{$\Delta_{\text{LHC$+$ILC}}$} & 
\multicolumn{2}{l}{$\Delta_{\text{LHC only}}$}\\ \hline
$\tan \beta$   &   10.000 &   10.000 & 0.036 &  1.3\\
$M_0$ (GeV)    &   70.000 &   70.000 & 0.070 &  1.4\\
$M_{1/2}$ (GeV)&  250.000 &  250.000 & 0.065 &  1.0\\
$A_0$ (GeV)    & -300.0   & -300.0   & 2.5   & 16.6\\
\hline
\end{tabular}
\end{center}
\caption{Fit results for fits within mSUGRA scenario. The meanings of
the columns are (starting from the left): SPS1a' values, the fitted mSUGRA
parameters, parameter uncertainties for a fit to LHC$+$ILC observables 
and parameter uncertainties for a fit to ``LHC only'' 
observables. In both cases theoretical uncertainties are not included.}
\label{tab:mSUGRAFitResults}
\end{table}

An important test is, however, to determine how well the experimental
results can be used to determine deviations from a specific high-scale
model such as mSUGRA. Therefore the mSUGRA high-energy parameters have
been fitted to the observables from the SPS1a' inspired scenario used
in Section~\ref{sec:SPS1aPrimeFit}, which is only slightly modified to
obtain exact unification in the first two generations of squarks and
sleptions. For this case, the total $\chi^2$s from the two fits
indicate that the LHC$+$ILC measurements are precise enough to
discriminate the modified scenario from exact mSUGRA ($\chi^2 =
3600$, 144 n.d.f.), while the LHC measurements
alone do not allow this insight ($\chi^2 = 0.5$, 25 n.d.f.).

\section{Conclusions}\label{sec:conclusions}

The Lagrangian parameters of the MSSM assuming universality for the
first and second generation and real parameters but without
assumptions on the SUSY breaking mechanism are correctly reconstructed
without usage of {\it a priori} information. This has been achieved
using precision measurements at the LHC and ILC as input to a global
fit exploiting the techniques implemented in the program Fittino.

Most of the Lagrangian parameters can be determined to a precision
around the percent level. For some parameters an accuracy of better
than 1~per mil is achievable if theoretical uncertainties are
neglected. As a (very) conservative estimate we have re-done the fit
using today's theoretical uncertainties. As expected it turns out that
they can significantly deteriorate the precision of the Lagrangian
parameter determination. More work is needed to improve the accurary
of theoretical predictions in order to fully benefit from the
experimental precision. The only parameter which cannot be strongly
constrained by the observables used in the presented fit is
$X_{\text{b}}$ resulting in a precision of only 25~\% for this
parameter which translates into an uncertainty of 130~\% on
$A_{\text{b}}$.

The strong (inter-sector) correlations between the fitted parameters
underline the importance of performing a global fit including all
sectors of the theory. As shown, fits in parameter subspaces can lead
to significant biases and wrong uncertainty estimates. 

The fitted parameters at the electroweak scale have been extrapolated
to high scales using the bottom-up approach taking into account all
correlations between the errors.  We have found that the gaugino
sector as well as the slepton sector can be reconstructed rather
precisely allowing for stringent tests of the underlying model in
these sectors. The accuracy deteriorates considerably for squark mass
parameters and the Higgs mass parameters.  The results clearly show
how far the squark mass parameters can deviate from the universality
assumption indicated by the measurements in the slepton sector. We
have also performed a top-down fit of the high-energy mSUGRA
parameters to combined LHC$+$ILC observables. Here we have found that
they are very sensitive to even tiny deviations from the mSUGRA
hypothesis. The discriminative power of ``LHC only'' inputs is
significantly weaker.

\section*{Acknowledgements}

The authors are grateful to Gudrid Moortgat-Pick, Georg Weiglein and
the whole SPA working group for very fruitful discussions.  W.~P.~is
supported by a MCyT Ramon y Cajal contract, by the Spanish grant
BFM2002-00345, by the European Commission Human Potential Program RTN
network HPRN-CT-2000-00148 and partly by the Swiss 'Nationalfonds'.

\clearpage
\newpage
\begin{appendix}
\section{List of Input Observables to the SPS1a' Inspired Fit}
\label{sec:InputsSPS1aPrimeAppendix}

\begin{table}[h]
\begin{center}
\begin{tabular}{l d{8} d{8} }
\hline
Observable & \multicolumn{1}{c}{Value} & \multicolumn{1}{c}{Exp.~uncertainty}  \\
\hline
   $ m_{\text{W}}$                 &  80.3371 \;\text{GeV} &  0.039  \;\text{GeV}                    \\
   $ m_{\text{Z}}$                 &  91.1187 \;\text{GeV} &  0.0021 \;\text{GeV}                    \\
   $ m_{\text{t}}$               & 178.0    \;\text{GeV} &  0.05    \;\text{GeV}  \\
   $ m_{\text{h}}$                 & 112.888  \;\text{GeV} &  0.05   \;\text{GeV} \\
   $ m_{\text{A}_{\text{pole}}}$                 & 374.228  \;\text{GeV} &  1.3    \;\text{GeV}  \\
   $ m_{\text{H}}$                 & 374.464  \;\text{GeV} &  1.3    \;\text{GeV} \\
   $ m_{\text{H}^{\pm}}$           & 383.131  \;\text{GeV} &  1.1    \;\text{GeV} \\
   $ m_{\tilde{q}_L}$              & 561.539   \;\text{GeV} &  9.8    \;\text{GeV} \\
   $ m_{\tilde{q}_R}$              & 543.35  \;\text{GeV} & 11.0    \;\text{GeV}  \\
   $ m_{\tilde{b}_1}$              & 502.059  \;\text{GeV} &  5.7    \;\text{GeV}  \\
   $ m_{\tilde{b}_2}$              & 541.81  \;\text{GeV} &  6.2    \;\text{GeV}  \\
   $ m_{\tilde{t}_1}$              & 365.819  \;\text{GeV} &  2.0    \;\text{GeV} \\
   $ m_{\tilde{e}_L}$              & 190.209  \;\text{GeV} &  0.2    \;\text{GeV} \\
   $ m_{\tilde{e}_R}$              & 124.883  \;\text{GeV} &  0.05   \;\text{GeV} \\
   $ m_{\tilde{\mu}_L}$            & 190.237  \;\text{GeV} &  0.5    \;\text{GeV} \\
   $ m_{\tilde{\mu}_R}$            & 124.837  \;\text{GeV} &  0.2    \;\text{GeV}  \\
   $ m_{\tilde{\tau}_1}$           & 107.292  \;\text{GeV} &  0.3    \;\text{GeV}  \\
   $ m_{\tilde{\tau}_2}$           & 195.290  \;\text{GeV} &  1.1    \;\text{GeV}  \\
   $ m_{\tilde{\text{g}}}$         & 603.639  \;\text{GeV} &  6.4    \;\text{GeV}  \\
   $ m_{\tilde{\chi}^0_1}$         & 97.7662  \;\text{GeV} &  0.05   \;\text{GeV} \\
   $ m_{\tilde{\chi}^0_2}$         & 184.345  \;\text{GeV} &  0.08   \;\text{GeV}  \\
   $ m_{\tilde{\chi}^0_3}$         & 404.134  \;\text{GeV} &  4.0    \;\text{GeV}  \\
   $ m_{\tilde{\chi}^0_4}$         & 417.037  \;\text{GeV} &  2.3    \;\text{GeV} \\
   $ m_{\tilde{\chi}^{\pm}_1}$     & 184.132  \;\text{GeV} &  0.55   \;\text{GeV} \\
   $ m_{\tilde{\chi}^{\pm}_2}$     & 418.495  \;\text{GeV} &  3.0    \;\text{GeV} \\
   Edge 3 with $ m_{\tilde{\chi}^0_1}$, $m_{\tilde{q}_L}$, $m_{\tilde{\chi}^0_2}$ & 449.679 \;\text{GeV} &  4.9 \;\text{GeV}\\
   Edge 3 with $m_{\tilde{\mu}_R}$, $m_{\tilde{q}_L}$, $m_{\tilde{\chi}^0_2}$ & 390.285 \;\text{GeV} & 3.35 \;\text{GeV}\\
   Edge 4 with $ m_{\tilde{\chi}^0_1}$, $m_{\tilde{\chi}^0_2}$, $m_{\tilde{\mu}_R}$, $m_{\tilde{q}_L}$ & 329.831 \;\text{GeV} & 4.2 \;\text{GeV}\\
   Edge 5 with $ m_{\tilde{\chi}^0_1}$, $m_{\tilde{\chi}^0_2}$, $m_{\tilde{\mu}_R}$, $m_{\tilde{q}_L}$ & 218.529 \;\text{GeV} & 3.44 \;\text{GeV} \\
\hline
\end{tabular}
\end{center}
\caption{Simulated measurements at LHC and at ILC running at 400 GeV, 500 GeV
      and 1 TeV center-of-mass energy. The
      values of the observables are taken from the prediction of SPheno version 2.2.2 for
      the SPS1a' inspired scenario. The experimental uncertainties on the masses
      are taken from~\cite{ref:LHCILC}. The definition of the edge numbers can be found in
      \cite{ref:FittinoProgram}.}
\label{tab:InputsSPS1aPrime}
\end{table}

\begin{table}[h]
\begin{center}
\begin{tabular}{l d{8} }
\hline
Observable & \multicolumn{1}{c}{Theor.~uncertainty} \\
\hline
   $ m_{\text{t}}$  & 0.1 \;\text{GeV}  \\
   $ m_{\text{h}}$  & 1.3 \;\text{GeV} \\
   $ m_{\text{A}_{\text{pole}}}$  & 0.7 \;\text{GeV} \\
   $ m_{\text{H}}$ &  0.7 \;\text{GeV} \\
   $ m_{\text{H}^{\pm}}$  &  0.7 \;\text{GeV} \\
   $ m_{\tilde{q}_L}$ & 10.2 \;\text{GeV} \\
   $ m_{\tilde{q}_R}$ &  9.4 \;\text{GeV} \\
   $ m_{\tilde{b}_1}$  &  8.0 \;\text{GeV} \\
   $ m_{\tilde{b}_2}$ &  10.2 \;\text{GeV} \\
   $ m_{\tilde{t}_1}$ &  5.4 \;\text{GeV} \\
   $ m_{\tilde{e}_L}$ &  0.4 \;\text{GeV} \\
   $ m_{\tilde{e}_R}$ &  1.2 \;\text{GeV} \\
   $ m_{\tilde{\mu}_L}$  &  0.4 \;\text{GeV} \\
   $ m_{\tilde{\mu}_R}$ &  1.2 \;\text{GeV} \\
   $ m_{\tilde{\tau}_1}$  &  0.5 \;\text{GeV} \\
   $ m_{\tilde{\tau}_2}$ &  0.5 \;\text{GeV} \\
   $ m_{\tilde{\text{g}}}$ &  1.4 \;\text{GeV} \\
   $ m_{\tilde{\chi}^0_1}$ &  0.4 \;\text{GeV} \\
   $ m_{\tilde{\chi}^0_2}$ &  1.2 \;\text{GeV} \\
   $ m_{\tilde{\chi}^0_3}$ &  1.2 \;\text{GeV} \\
   $ m_{\tilde{\chi}^0_4}$ &  1.2 \;\text{GeV} \\
   $ m_{\tilde{\chi}^{\pm}_1}$ &  1.3 \;\text{GeV} \\
   $ m_{\tilde{\chi}^{\pm}_2}$  &  1.3 \;\text{GeV} \\
   Edge 3 with $ m_{\tilde{\chi}^0_1}$, $m_{\tilde{q}_L}$, $m_{\tilde{\chi}^0_2}$  & 4.5 \;\text{GeV}\\
   Edge 3 with $m_{\tilde{\mu}_R}$, $m_{\tilde{q}_L}$, $m_{\tilde{\chi}^0_2}$ & 3.9 \;\text{GeV}\\
   Edge 4 with $ m_{\tilde{\chi}^0_1}$, $m_{\tilde{\chi}^0_2}$, $m_{\tilde{\mu}_R}$, $m_{\tilde{q}_L}$ & 3.3 \;\text{GeV}\\
   Edge 5 with $ m_{\tilde{\chi}^0_1}$, $m_{\tilde{\chi}^0_2}$, $m_{\tilde{\mu}_R}$, $m_{\tilde{q}_L}$ & 2.19 \;\text{GeV}\\
\hline
\end{tabular}
\end{center}
\caption{Theoretical uncertainties on the masses (taken from \cite{ref:SPAPaper})
  and the induced ones on the edge variables.}
\label{tab:TheoInputsSPS1aPrime}
\end{table}

\clearpage
\newpage
\section{Correlation Matrix of the SPS1a' Inspired Fit}
\label{sec:CorrelationMatrixAppendix}

Table~\ref{tab:CorrelationMatrix1} and \ref{tab:CorrelationMatrix2}
show the full correlation matrix from the fit for the SPS1a' inspired
scenario described in Section~\ref{sec:SPS1aPrimeFit}.

\begin{table}[h]
{\scriptsize
\begin{center}
\begin{tabular}{l d{3} d{3} d{3} d{3} d{3} d{3} d{3} d{3} d{3} d{3}}
\hline
Parameter & \multicolumn{1}{c}{$\tan \beta$} & \multicolumn{1}{c}{$\mu$} & \multicolumn{1}{c}{$X_{\tau}$}
& \multicolumn{1}{c}{$M_{\tilde{e}_R}$} & \multicolumn{1}{c}{$M_{\tilde{\tau}_R}$}
& \multicolumn{1}{c}{$M_{\tilde{e}_L}$}   & \multicolumn{1}{c}{$M_{\tilde{\tau}_L}$}
& \multicolumn{1}{c}{$X_{\text{t}}$} & \multicolumn{1}{c}{$X_{\text{b}}$} & \multicolumn{1}{c}{$M_{\tilde{q}_R}$}\\
 \hline
$\tan \beta$                &  1.000 & -0.509 & -0.238 &  0.042 &  0.032 & -0.125 & -0.321 &  0.030 & -0.569 &  0.033 \\
$\mu$                       & -0.509 &  1.000 &  0.014 & -0.038 & -0.028 &  0.303 &  0.308 & -0.064 &  0.151 &  0.031 \\
$X_{\tau}$                  & -0.238 &  0.014 &  1.000 &  0.055 & -0.431 & -0.076 &  0.225 & -0.137 & -0.091 & -0.058 \\
$M_{\tilde{e}_R}$           &  0.042 & -0.038 &  0.055 &  1.000 &  0.369 & -0.631 & -0.329 &  0.026 & -0.109 & -0.793 \\
$M_{\tilde{\tau}_R}$        &  0.032 & -0.028 & -0.431 &  0.369 &  1.000 & -0.254 & -0.339 &  0.062 &  0.074 & -0.362 \\
$M_{\tilde{e}_L}$           & -0.125 &  0.303 & -0.076 & -0.631 & -0.254 &  1.000 &  0.368 & -0.048 &  0.112 &  0.557 \\
$M_{\tilde{\tau}_L}$        & -0.321 &  0.308 &  0.225 & -0.329 & -0.339 &  0.368 &  1.000 & -0.023 &  0.121 &  0.320 \\
$X_{\text{t}}$              &  0.030 & -0.064 & -0.137 &  0.026 &  0.062 & -0.048 & -0.023 &  1.000 & -0.360 &  0.097 \\
$X_{\text{b}}$              & -0.569 &  0.151 & -0.091 & -0.109 &  0.074 &  0.112 &  0.121 & -0.360 &  1.000 & -0.090 \\
$M_{\tilde{q}_R}$           &  0.033 &  0.031 & -0.058 & -0.793 & -0.362 &  0.557 &  0.320 &  0.097 & -0.090 &  1.000 \\
$M_{\tilde{b}_R}$           & -0.049 &  0.037 & -0.016 &  0.193 &  0.080 & -0.133 & -0.011 &  0.076 &  0.004 &  0.091 \\
$M_{\tilde{t}_R}$           & -0.166 &  0.125 & -0.011 & -0.335 & -0.141 &  0.327 &  0.210 & -0.371 &  0.307 &  0.082 \\
$M_{\tilde{q}_L}$           & -0.065 &  0.025 & -0.035 &  0.192 &  0.069 & -0.084 & -0.001 &  0.144 & -0.014 &  0.076 \\
$M_{\tilde{t}_L}$           & -0.005 & -0.094 &  0.054 &  0.109 &  0.001 & -0.138 & -0.066 & -0.270 &  0.092 &  0.070 \\
$M_1$                       & -0.147 & -0.263 &  0.118 &  0.169 &  0.051 & -0.177 &  0.032 &  0.044 &  0.130 & -0.244 \\
$M_2$                       &  0.229 & -0.694 & -0.299 & -0.047 &  0.062 &  0.087 &  0.024 &  0.128 &  0.042 & -0.058 \\
$M_3$                       &  0.044 &  0.003 & -0.036 &  0.078 &  0.065 & -0.046 & -0.061 &  0.076 & -0.070 & -0.180 \\
$m_{\text{A}_{\text{run}}}$ & -0.594 &  0.143 & -0.048 & -0.108 &  0.056 &  0.110 &  0.130 & -0.223 &  0.972 & -0.076 \\
$m_{\text{t}}$              & -0.004 & -0.013 &  0.004 & -0.005 &  0.018 & -0.018 &  0.039 &  0.186 & -0.029 &  0.011 \\
\hline
\end{tabular}
\end{center}
}
\caption{Correlation matrix of the Fittino SPS1a' motivated fit, part I.\vspace{2mm}}
\label{tab:CorrelationMatrix1}
\end{table}

\begin{table}[h]
{\scriptsize
\begin{center}
\begin{tabular}{l d{3} d{3} d{3} d{3} d{3} d{3} d{3} d{3} d{3} d{3}}
\hline
Parameter & \multicolumn{1}{c}{$M_{\tilde{b}_R}$} & \multicolumn{1}{c}{$M_{\tilde{t}_R}$}
& \multicolumn{1}{c}{$M_{\tilde{q}_L}$} & \multicolumn{1}{c}{$M_{\tilde{t}_L}$}
& \multicolumn{1}{c}{$M_1$}   & \multicolumn{1}{c}{$M_2$}
& \multicolumn{1}{c}{$M_3$} & \multicolumn{1}{c}{$m_{\text{A}_{\text{run}}}$} & \multicolumn{1}{c}{$m_{\text{t}}$} \\
 \hline
$\tan \beta$                 & -0.049 & -0.166 & -0.065 & -0.005 & -0.147 &  0.229 &  0.044 & -0.594 & -0.004 \\
$\mu$                        &  0.037 &  0.125 &  0.025 & -0.094 & -0.263 & -0.694 &  0.003 &  0.143 & -0.013 \\
$X_{\tau}$                   & -0.016 & -0.011 & -0.035 &  0.054 &  0.118 & -0.299 & -0.036 & -0.048 &  0.004 \\
$M_{\tilde{e}_R}$            &  0.193 & -0.335 &  0.192 &  0.109 &  0.169 & -0.047 &  0.078 & -0.108 & -0.005 \\
$M_{\tilde{\tau}_R}$         &  0.080 & -0.141 &  0.069 &  0.001 &  0.051 &  0.062 &  0.065 &  0.056 &  0.018 \\
$M_{\tilde{e}_L}$            & -0.133 &  0.327 & -0.084 & -0.138 & -0.177 &  0.087 & -0.046 &  0.110 & -0.018 \\
$M_{\tilde{\tau}_L}$         & -0.011 &  0.210 & -0.001 & -0.066 &  0.032 &  0.024 & -0.061 &  0.130 &  0.039 \\
$X_{\text{t}}$               &  0.076 & -0.371 &  0.144 & -0.270 &  0.044 &  0.128 &  0.076 & -0.223 &  0.186 \\
$X_{\text{b}}$               &  0.004 &  0.307 & -0.014 &  0.092 &  0.130 &  0.042 & -0.070 &  0.972 & -0.029 \\
$M_{\tilde{q}_R}$            &  0.091 &  0.082 &  0.076 &  0.070 & -0.244 & -0.058 & -0.180 & -0.076 &  0.011 \\
$M_{\tilde{b}_R}$            &  1.000 &  0.009 & -0.017 & -0.077 & -0.034 &  0.024 & -0.105 &  0.026 &  0.012 \\
$M_{\tilde{t}_R}$            &  0.009 &  1.000 & -0.036 & -0.428 & -0.048 &  0.045 & -0.236 &  0.249 & -0.063 \\
$M_{\tilde{q}_L}$            & -0.017 & -0.036 &  1.000 &  0.057 &  0.006 & -0.200 & -0.275 &  0.011 &  0.017 \\
$M_{\tilde{t}_L}$            & -0.077 & -0.428 &  0.057 &  1.000 & -0.002 & -0.140 & -0.225 &  0.106 & -0.081 \\
$M_1$                        & -0.034 & -0.048 &  0.006 & -0.002 &  1.000 &  0.293 &  0.055 &  0.141 & -0.060 \\
$M_2$                        &  0.024 &  0.045 & -0.200 & -0.140 &  0.293 &  1.000 &  0.109 &  0.040 & -0.005 \\
$M_3$                        & -0.105 & -0.236 & -0.275 & -0.225 &  0.055 &  0.109 &  1.000 & -0.086 &  0.008 \\
$m_{\text{A}_{\text{run}}}$  &  0.026 &  0.249 &  0.011 &  0.106 &  0.141 &  0.040 & -0.086 &  1.000 & -0.008 \\
$m_{\text{t}}$               &  0.012 & -0.063 &  0.017 & -0.081 & -0.060 & -0.005 &  0.008 & -0.008 &  1.000 \\
\hline
\end{tabular}
\end{center}
}
\caption{Correlation matrix of the Fittino SPS1a' motivated fit, part II.\vspace{2mm}}
\label{tab:CorrelationMatrix2}
\end{table}


\end{appendix}


\end{document}